\documentclass[USenglish,twocolumn]{article}

\ifx\directlua\undefined\ifx\XeTeXcharclass\undefined
  \usepackage[utf8]{inputenc}                           %
  \else\RequirePackage[no-math]{fontspec}[2017/03/31]\fi %
  \else\RequirePackage[no-math]{fontspec}[2017/03/31]\fi %
\usepackage[big,online]{dgruyter}

\theoremstyle{dgthm}

\theoremstyle{dgdef}

\usepackage[dvipsnames,table,xcdraw]{xcolor}
\definecolor{apsblue}{HTML}{2e3092}
\hypersetup{
    colorlinks=true,
	citecolor=apsblue,
	linkcolor=apsblue,
	filecolor=magenta,
	urlcolor=apsblue,
}

\usepackage[
    backend=biber,
    style=ieee,
    sorting=none
]{biblatex}
\usepackage{physics}
\newcommand{\iu}{\mathrm{i}\mkern1mu}
\newcommand{\eu}{\mathrm{e}\mkern1mu}
\newcommand{\bsans}[1]{\textbf{\textsf{#1}}}

\usepackage[dvipsnames]{xcolor}

\begin{document}

\articletype{Research Article}

\author[1]{Ivan Toftul$^{\ddagger}$}
\author[2]{Dhruv Hariharan$^{\ddagger}$}
\author[3]{Pavel Tonkaev$^{\ddagger}$}
\author[4]{Fangxing Lai}
\author[5]{Qinghai Song}
\author*[6]{Yuri Kivshar}

\affil[1]{Research School of Physics, Australian National University, Canberra ACT 2601, Australia, ivan.toftul@anu.edu.au; 0000-0003-3588-5403}
\affil[2]{Research School of Physics, Australian National University, Canberra ACT 2601, Australia, 0009-0003-4610-7014}
\affil[3]{Research School of Physics, Australian National University, Canberra ACT 2601, Australia, 0000-0003-1849-0653}
\affil[4]{Ministry of Industry and Information Technology Key Lab of Micro-Nano Optoelectronic Information System,
Guangdong Provincial Key Laboratory of Semiconductor Optoelectronic Materials and Intelligent Photonic Systems,
Harbin Institute of Technology, Shenzhen 518055, People’s Republic of China; 0000-0001-9176-0205}
\affil[5]{Ministry of Industry and Information Technology Key Lab of Micro-Nano Optoelectronic Information System,
Guangdong Provincial Key Laboratory of Semiconductor Optoelectronic Materials and Intelligent Photonic Systems,
Harbin Institute of Technology, Shenzhen 518055, People’s Republic of China; 0000-0003-1048-411X}
\affil[6]{Research School of Physics, Australian National University, Canberra ACT 2601, Australia, yuri.kivshar@anu.edu.au; 0000-0002-3410-812X}

\title{Monoclinic nonlinear metasurfaces for resonant engineering of polarization states}

\runningtitle{Monoclinic nonlinear metasurfaces \dots }
\runningauthor{I. Toftul$^{\ddagger}$, D. Hariharan$^{\ddagger}$, P. Tonkaev$^{\ddagger}$, F. Lai, Q. Song, Y. Kivshar}

\abstract{
Polarization is a fundamental property of light that can be engineered and controlled efficiently with optical metasurfaces. Here, we employ {\it chiral metasurfaces} with monoclinic lattice geometry and achiral meta-atoms for resonant engineering of polarization states of light. We demonstrate, both theoretically and experimentally, that a monoclinic metasurface can convert linearly polarized light into elliptically polarized light not only in the linear regime but also in the nonlinear regime with the resonant generation of the third-harmonic field. We reveal that the ellipticity of the fundamental and higher-harmonic fields depends critically on the angle of the input linear polarization, and the effective chiral response of a monoclinic lattice plays a significant role in the polarization conversion.
}


\maketitle

\section{Introduction}

Among many remarkable achievements associated with the name Federico Capasso, metasurfaces {\it} play an important role as efficient planar components of future photonic devices~\cite{roadmap,roadmap2,roadmap3}.  
One of the fundamental functionalities of metasurfaces is the control of polarization of light, which is a versatile degree of freedom that can be manipulated or engineered for numerous applications~\cite{Goldstein2003,Collett2005}. Conventional approaches to manipulating polarization often rely on bulk components such as wave plates. Although effective, these components are constrained by their size.

\begin{figure}
    \centering
    \includegraphics[width=0.8\linewidth]{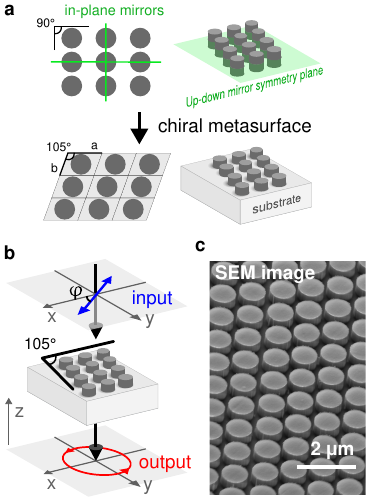}
    \caption{\textbf{Concept of monoclinic chiral metasurfaces.} 
    \bsans{a} Design principle of the chiral metasurfaces utilizing monoclinic meta-atom lattices.
    \bsans{b} Schematic of a chiral metasurface which transforms incident linearly polarized plane wave into elliptically polarized light. 
    \bsans{c} Scanning electron microscope (SEM) image of the fabricated metasurface. }
    \label{fig:concept}
\end{figure}

Metasurfaces, composed of subwavelength structures, that is, meta-atoms, are powerful tools for overcoming size and weight limitations, allowing ultra-thin and multifunctional optical devices~\cite{Jung2021ChemRev,Chen2020AdvMater,Wen2018AOM}.
The operational principle for using metasurfaces in the control of polarization relies on transforming an incident waveform into an ensemble of individual beams generated by meta-atoms with different polarization states that beat along the optical axis, thereby changing the resulting polarization at will. Many recent papers have been devoted to the study of linear polarization conversion with single metasurfaces (see, e.g, Refs. ~\cite{linear00,linear0,shen_lin,linear1,linear2}), as well as multiple metasurfaces with stacking and twisted configurations~\cite{linear3,linear4}. Polarization transformations with metasurfaces have been extensively explored by the Federico Capasso group
~\cite{capasso0,capasso,capasso2}. In particular, the designer-specified polarization response was shown to be employed for computer-generated holograms whose far-fields implement parallel polarization analysis and customized waveplates~\cite{capasso2}. Additionally, full-stokes polarization encoding in metasurfaces has been demonstrated in both the near ~\cite{asadulina2024all, polevoy2023excitation} and far-field regimes ~\cite{zuo2023chip, deng2024poincare}. 

\textit{Chiral metasurfaces} are particularly suitable for polarization engineering because of their inherit ability to mix polarization states, e.g. to convert directly linearly polarized light into elliptically or circularly polarized states~\cite{Overig2021APS}. The use of resonant effects in metasurfaces has been demonstrated to enhance polarization conversion efficiencies and enable functionalities such as circularly polarized lasing and high-contrast polarization detection~\cite{Katsantonis2024ACSPhot,Bai2021ACSNano, Zhang2022Science} as well as achieving huge imbalance in the third harmonic intensity depending on the helicity of the input field~\cite{Gandolfi2021PRA}. 

In this work, we uncover the hidden potential of resonant chiral monoclinic metasurfaces, recently introduced and characterized in Ref.\cite{Toftul2024PRL}, for polarization conversion. While the previous study focused on circularly polarized input, this work explores the metasurfaces' response from linearly polarized input, and expands this concept to the nonlinear polarization conversion for the generation of third-harmonic chiral fields (Fig.~\ref{fig:concept}). 
Using both computational and experimental approaches, we demonstrate their ability to convert linearly polarized light into elliptically polarized light in both the linear and nonlinear regimes. Specifically, we explore the third harmonic generation (THG) process and show that the ellipticity of the generated light is strongly dependent on the input polarization angle. By analyzing the role of chiral resonances, we underpin the underlying mechanisms that govern these effects, highlighting the versatility of monoclinic metasurfaces as compact polarization engineering platforms.
This builds on previous work exploring multifunctional metasurfaces for polarization conversion and control~\cite{Deng2021ACSNano,Mustafa2018SciRep}.

\section{Results}

The object of our research is a chiral resonant dielectric metasurface, which consists of Si cylinders in a monoclinic arrangement on a SiO$_2$ substrate. 
Such a metasurface is geometrically chiral, i.e. it does not possess any mirror symmetry: the up-down mirror symmetry is broken by the substrate, and all in-plane mirror symmetries are broken by the monoclinic arrangement itself. 
Circular dichroism studies of this metasurface can be found in Ref.~\cite{Toftul2024PRL}. 

\begin{figure}
    \centering
    \includegraphics[width=0.7\linewidth]{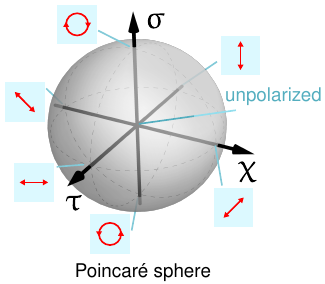}
    \caption{Poincar\`e sphere and illustration of the polarization states.}
    \label{fig:poincare}
\end{figure}

Realization of a polarization transformation can be conveniently visualized on the unit Poincar\'e sphere~\cite{Collett2005,Cisowski2022RMP} via the following three parameters:
\begin{align}
    \tau &= |e_x|^2 - |e_y|^2  = \frac{I_{\text{V}} - I_{\text{H}}}{I_{\text{V}} + I_{\text{H}}} \nonumber \\
    \chi &= 2 \Re\left( e_x^* e_y\right) = \frac{I_{\text{D}} - I_{\text{A}}}{I_{\text{D}} + I_{\text{A}}}  \label{eq:Stokes} \\
    \sigma &= 2 \Im\left( e_x^* e_y\right) = \frac{I_{\text{L}} - I_{\text{R}}}{I_{\text{L}} + I_{\text{R}}}. \nonumber
\end{align}
These parameters show the degrees of the vertical/horizontal ($\tau$), diagonal/anti-diagonal ($\chi$), and right-hand/left-hand circular polarizations ($\sigma$), see Fig.~\ref{fig:poincare}. 
Here, the complex components of the electric field $e_x$ and $e_y$ are normalized so that $|e_x|^2 + |e_y|^2 = 1$. 
In the far-field plane the field is transverse, i.e. $e_z = 0$.
Experimentally, Stokes parameters are calculated by performing six different intensity measurements: vertical and horizontal $I_{\text{V}/\text{H}}$,  diagonal and anti-diagonal $I_{\text{D}/\text{A}}$, polarized on the right and left circular $I_{\text{R}/\text{L}}$. 
The normalized Stokes parameters satisfy $\tau^2 + \chi^2 + \sigma^2 \leq  1$, where ``$<$'' is achieved for the partially or fully unpolarized signal.

\begin{figure*}
    \centering
    \includegraphics[width=0.85\linewidth]{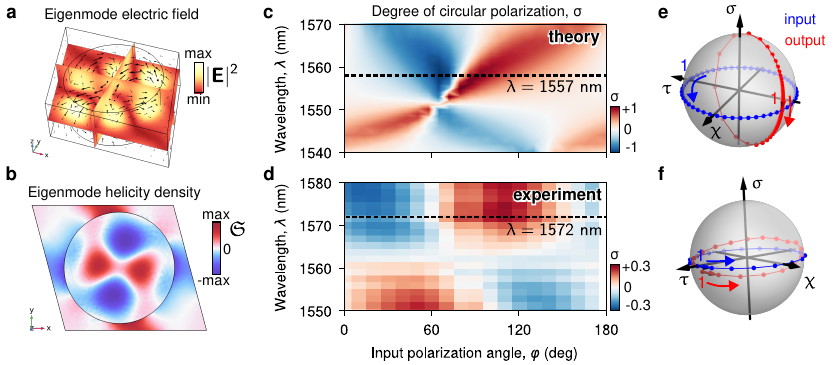}
    \caption{ \textbf{Linear chiral optical properties}. Theoretical electric field (\bsans{a}) and helicity density (\bsans{b}) of the excited eigenmode at $1557\text{ nm}$. Theoretical (\bsans{c}) and experimental (\bsans{d}) degree of circular polarization $\sigma$ dependence on input polarization angle and wavelength. Output theoretical (\bsans{e}) and experimental (\bsans{f}) polarization state (red dots) for various input linear polarization angles (blue dots) plotted on the Poincare sphere.}
    \label{fig:linear}
\end{figure*}

The metasurface exhibits resonant behavior. The design parameters are as follows: the meta-atoms are cylinders with a height of $400$~nm and a radius of $430$~nm, composed of silicon (Si); the monoclinic arrangement is defined by two lattice vectors with lengths $a=1100$~nm and $b=1000$~nm, and a lattice angle of $105^{\circ}$ (Fig.~\ref{fig:concept}\bsans{a}); the substrate is made of SiO$_2$. We find several eigenmodes of the systems in the vicinity of the wavelength telecommunication range. In particular, there is an eigenmode with a resonant wavelength of $1553$~nm and $Q$-factor of $Q \approx 100$. We show its electric field distribution in Fig.~\ref{fig:linear}\bsans{a}. 
Geometrically chiral metasurface supports chiral eigenmodes, however, their high level of chirality --- selective interaction with right and left circularly polarized output channels --- is not guaranteed, and has to be engineered~\cite{Gorkunov2020PRL,Shalin2023,Gorkunov2024AOM,Toftul2024PRL}. 
It was shown in Ref.~\cite{Toftul2024PRL} that the eigenmode in Fig.~\ref{fig:linear}\bsans{a} exhibits strong chiral properties, which manifests in a strong circular dichroism.

Additionally to the transmission properties of the resonant chiral metasruface with monoclinic lattice arrangement, we also shed the light on its  \textit{local} properties.
Helicity density is a local property of the electromagnetic field distribution. In particular it has direct application in chiral sensing~\cite{Solomon2020ACR,Toftul2024arXivComplexForces}, which is often referred as one of the promising applications of the chiral metasurfaces. 
For monochromatic field at frequency $\omega$ it is written as~\cite{Bliokh2011PRA,Bliokh2014PRL,Cameron2012}
\begin{equation}
    \mathfrak{S} = \frac{1}{2\omega c} \Im \left( \mathbf{H}^* \cdot \mathbf{E}\right),
    \label{eq:helicity}
\end{equation}
where $\mathbf{E}$ and $\mathbf{H}$ are the electric and magnetic fields, $c$ is the speed of sound. 
The quantity \eqref{eq:helicity} characterizes the difference between the numbers of right-hand and left-hand circularly polarized photons. 
We plot the distribution of helicity density of the eigenmode in Fig.~\ref{fig:linear}\bsans{b}.

Next, we study the polarization transformation of such metasurface for the linearly polarized input light in the linear and nonlinear regime, i.e. polarization state of the third harmonic generation signal.

\subsection{Polarization transformation in linear transmission}

Here we examine the manifestation of the chiral mode of choice in the context of polarization transformation for the linearly polarized input. 
We set the input field to be a monochromatic linearly polarized plane wave  at frequency $\omega$ and wave vector $\vb{k} = - \vu{z} k$ (we assume a $\eu^{- \iu \omega t}$ time dependence):
\begin{equation}
    \vb{E}_{\text{in}}^{(\omega)} = E_0 \left( \vu{x} \cos \varphi + \vu{y} \sin \varphi \right) \eu^{-\iu k z},
\end{equation}
where  angle $\varphi$ shows linear polarization orientation with respect to the $x$-axis, $\vu{x}$, $\vu{y}$, $\vu{z}$ are the Cartesian unit vectors. Based on Eq.~\eqref{eq:Stokes}, this implies the input Stokes parameters to be 
\begin{equation}
    \tau_{\text{in}} = \cos(2\varphi), \quad 
    \chi_{\text{in}} = \sin(2\varphi), \quad 
    \sigma_{\text{in}} = 0.
    \label{eq:stokes_input}
\end{equation}
We incrementally change polarization angle $\varphi$ in the range $[0, \pi]$ with a constant step $\Delta \varphi$, as the results are $\pi$-periodic. 

Adjusting the polarization angle $\varphi$ and the wavelength $\lambda$ of the incident wave, we examine the polarization of the outgoing signal (Fig.~\ref{fig:linear}\bsans{c}). Near the chiral resonance, the circular polarization degree $\sigma$ is strongly dependent on $\lambda$ and $\varphi$. At certain $\lambda$ and $\varphi$, the transmitted field approaches circular polarization, $|\sigma| \approx 1$. 
Although the input polarization states \eqref{eq:stokes_input} are evenly spaced, the output states are not. 
This is illustrated on the Poincar\`e sphere in Fig.~\ref{fig:linear}\bsans{e}. In particular, the output states form a circle on the sphere, indicated by red dots.

\begin{figure*}
    \centering
    \includegraphics[width=0.8\linewidth]{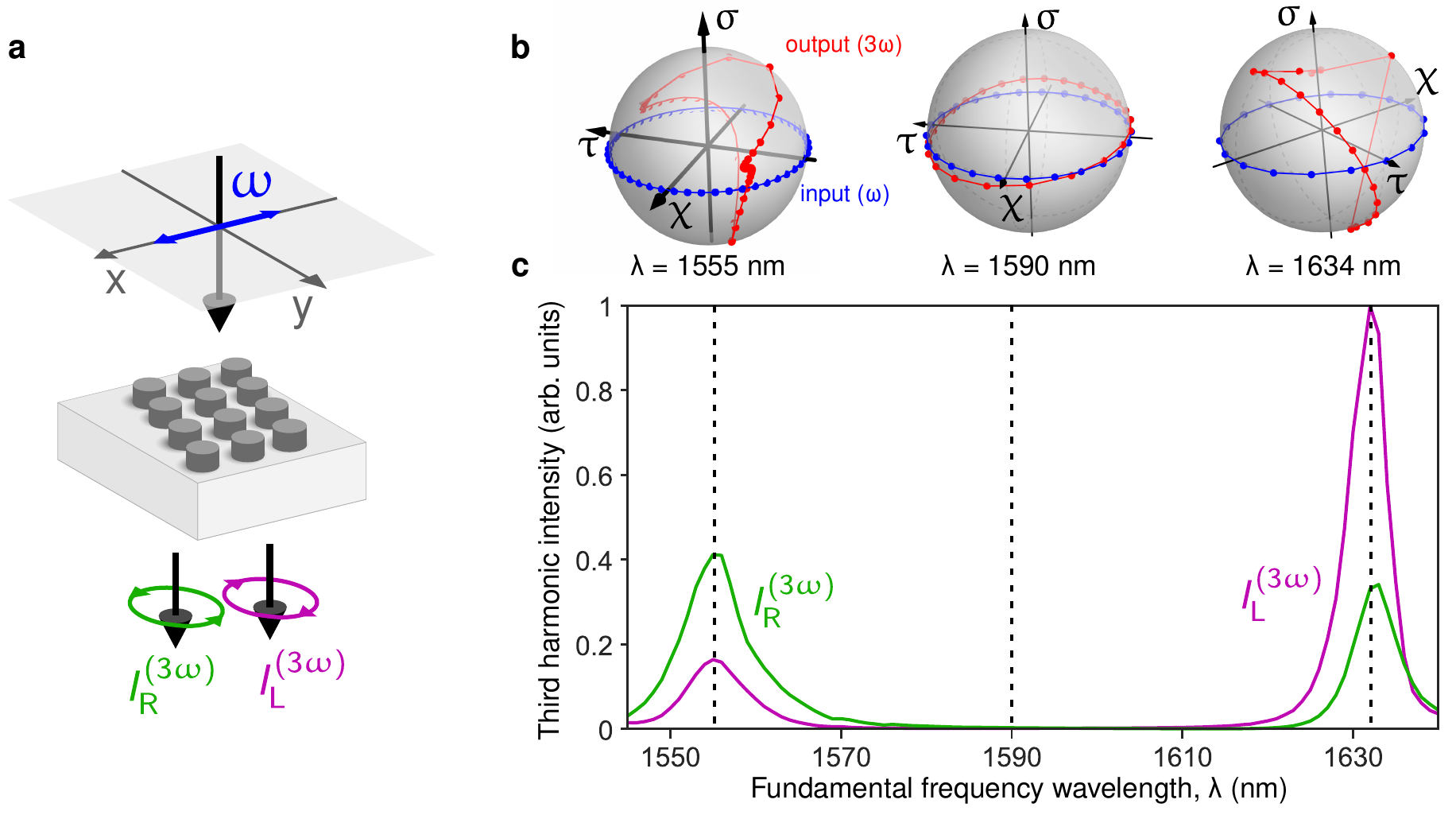}
    \caption{\textbf{Nonlinear theory.} 
    \bsans{a} Third harmonic intensity output from a linearly polarized beam along $x$-axis. \bsans{c} The green and magenta lines are the intensity of the right and left circular polarization intensities of the third harmonic signal, respectively. \bsans{b}  Poincar\`e spheres which show simulated polarization states of the third harmonic output taken at different wavelengths for linear polarization input at different angles on the fundamental frequency. }
    \label{fig:nonlinear_theory}
\end{figure*}

To validate our theoretical prediction regarding the polarization change, we measure all Stokes parameters of light transmitted through the metasurface taking into account Eq.~\eqref{eq:Stokes} (see the Methods section for experimental details). The experimentally determined values of $\sigma$, as a function of the wavelength and the input polarization angle, are presented in Fig.~\ref{fig:linear}\bsans{d}. The output light exhibits elliptical polarization around the optical resonance. The behavior observed qualitatively agrees with the theoretical protection. However, the degree of circular polarization varies from $-0.18$ to $0.26$ at $1572$~nm. The output polarization state is illustrated in Fig.~\ref{fig:linear}\bsans{f} by red dots, while the input polarization state is also depicted by blue dots. The experimental pattern is similar to the theoretical prediction. The difference between theoretical and experimental results is primarily attributed to fabrication imperfections and the broad range of incident light wave vectors in the experiment, which result in different mode excitations to the theoretical prediction. We provide theoretical and experimental results for wider range in Supplementary Materials.

\subsection{Polarization engineering of the third harmonic generation}

At higher intensities, the electron oscillations within a dielectric structure become \textit{anharmonic}, which can be effectively described using the extended Lorentz model~\cite{Boyd2008}. 
In its bulk crystalline form, silicon is a centrosymmetric material, making second harmonic generation (SHG) symmetry-forbidden (except surface effects ~\cite{Tonkaev2024PRR}). Therefore, the third harmonic generation (THG) was examined to demonstrate the nonlinear behavior of the metasurface. 
The nonlinear polarization current responsible for THG can be expressed as $\mathbf{P}^{(3\omega)} = \varepsilon_0 \hat{\chi}^{(3)} \mathbf{E}^{(\omega)} \mathbf{E}^{(\omega)} \mathbf{E}^{(\omega)}$, where $\varepsilon_0$ is the dielectric constant, $\mathbf{E}^{(\omega)}$ is the electric field at the fundamental frequency, and $\hat{\chi}^{(3)}$ is the fourth-rank nonlinear susceptibility tensor. Silicon has space group $m3m$, which results in only 21 nonzero elements in $\hat{\chi}^{(3)}$ with only 4 independent~\cite{Boyd2008}.
While it is possible to find values of each component experimentally in some approximations, their values are usually of the same order of magnitude~\cite{Zhang20105thInternationalSymposiumonAdvancedO,Moss1989OL}. 
For simplicity, we assume all non-zero components to be equal, e.g. as it is done in~\cite{Koshelev2023ACS}.
Moreover, in Supplementary Materials we show that even approximation of isotropic nonlinear response gives practically the same results.
To model THG, we employ the undepleted pump approximation and simulate the process in COMSOL Multiphysics using a two-step approach. In this framework, the nonlinear polarization $\mathbf{P}^{(3\omega)}$ is used as the initial condition for solving the higher harmonic wave equations.

The coupling strength of the incident field to the resonant mode strongly depends on the overlap integral between the two~\cite{Toftul2024PRL,Gorkunov2020PRL,Gorkunov2024AOM,Koshelev2023ACSphot}. This coupling shows a strong polarization dependence, providing different field distribution at the fundamental frequency $\mathbf{E}^{(\omega)}$ for different input polarizations, and hence different $\mathbf{P}^{(3\omega)}$ and third harmonic response.

\begin{figure*}
    \centering
    \includegraphics[width=\linewidth]{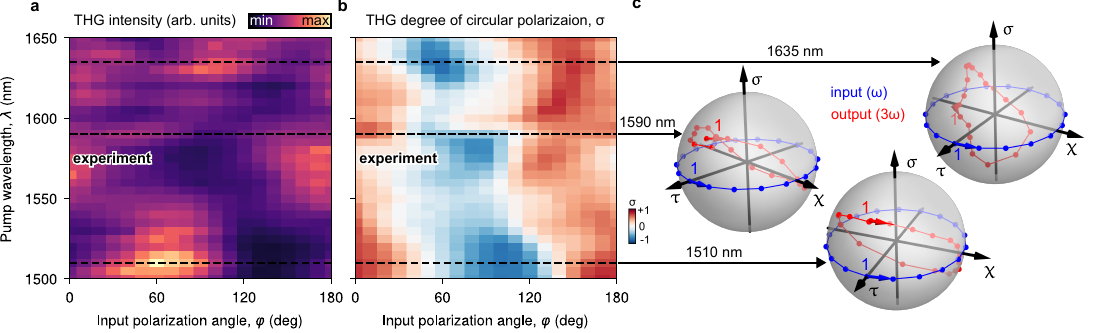}
    \caption{\textbf{Experimental nonlinear polarization properties.} THG intensity (\bsans{a}) and degree of circular polarization (\bsans{b}) from the metasurface as a function of the pump wavelength and polarization angle.  \bsans{c} Poincar\`e spheres illustrating polarization states of the output THG at $1510$~nm, $1590$~nm and $1635$~nm wavelengths for linear polarized input, varying linear polarization angle at the fundamental harmonic.}
    \label{fig:nonlinear_exp}
\end{figure*}

We evaluate the nonlinear numerical results by calculating the THG emitted along the zeroth order for a linearly polarized pump and decomposing THG transmission signal into the right- and left-circular polarized components, $I_{\text{R}}^{(3\omega)}$ and $I_{\text{L}}^{(3\omega)}$ (Fig.~\ref{fig:nonlinear_theory}\bsans{a}). The wavelength-dependent THG intensities for right- and left- circular polarized components are shown in Fig.~\ref{fig:nonlinear_theory}\bsans{c} by the green and magenta lines, respectively. The results exhibit a pronounced resonant behavior revealing significant enhancement of the THG at the vicinity of the structure resonances. Remarkably, for the first resonance ($1555$~nm) the right-circular polarized component dominates, while for the second resonance ($1634$~nm) the left-circular polarized component becomes dominant. Next, we investigate the effect of varying the polarization angle of the pump at specific wavelengths and extract all polarization parameters for the THG. The simulated polarization states are represented on the Poincar\`e spheres in Fig.~\ref{fig:nonlinear_theory}\bsans{b}. 
Near the chiral resonances, the polarization trajectories for pump wavelengths of $1555$~nm and $1634$~nm exhibit high degrees of circular polarization, with $\abs{\sigma}$ values approaching $1$. This indicates that highly circularly polarized THG is generated near these resonances. Unlike the linear case, the polarization points do not lie within a single plane and lack a clear pattern. This complexity likely results from the high density of resonant chiral states in the vicinity of $3\omega$ (see Supplementary Material), where each mode shows a slightly different coupling coefficient --- i.e., an overlap integral between the fundamental harmonic field and the high harmonic resonance --- that varies significantly with the linear polarization angle with the resonance at the fundamental harmonic, as the field intensity is less homogeneous. In contrast, for wavelengths away from the resonances (e.g., $1590$~nm, as shown in Fig. \ref{fig:nonlinear_theory}\bsans{b}), the output polarization closely mimics the input polarization, indicating the significance of the chiral resonance at the fundamental harmonic. This aspect requires further investigation and is not fully covered in the current paper.

To test the theoretical predictions, We measure the THG from the metasurface (see the Methods section for more details). The laser wavelength was tuned from $1500$--$1730$~nm at $5$~nm intervals, while the polarization angle was varied from $0^{\circ}$ to $180^{\circ}$ with $10^{\circ}$. For each combination of these parameters, we record the THG spectra and extract the maximum values. 
Fig.~\ref{fig:nonlinear_exp}\bsans{a} shows the maximum THG value as a function of the pump wavelength and input polarization angle. 
We observe THG enhancement in both expected and unexpected spectral regions. 
The exact reason of such discrepancy between numerical simulations and experimental observation is rather unknown. However we speculate here on possible reasons: 
\begin{enumerate}
    \item[(i)] imperfections in fabrications are much more noticeable on the scale of $\lambda^{(3\omega)}/n \simeq 125~\text{nm}$ (the typical wavelength of the TH signal inside metasurface material with refractive index $n$), specifically at the cylinders edges where the high-harmonic modes are mostly localized as they resemble whispery gallery modes;
    \item[(ii)] difference in the refractive index dispersion used in simulations and dispersion of the real sample, as the change of the refractive index tend to shift the position of the resonances; 
    \item[(iii)] deviation of excitation shape from a plane wave, which was used in the theoretical calculations. 
\end{enumerate}
Numerical simulations of the THG intensity signal of the range shown in Fig.~\ref{fig:nonlinear_exp}\bsans{a} and typical mode profiles at $3\omega$ frequencies are shown in Supplemental Materials.

To investigate the degree of circular polarization $\sigma$ we extract this parameter from the THG data taking into account Eq.~\eqref{eq:Stokes} --- the dependence on the pump wavelength and input polarization angle is shown in Fig.~\ref{fig:nonlinear_exp}\bsans{b}. 
The results reveal complex dependencies with significant changes in the polarization state as the pump polarization angle. We provide a possible justification in the Supplementary Material. We further extract the polarization parameters for the resonant wavelengths $1510$~nm and $1635$~nm. These values are plotted on Poincar\`e spheres, showing $\sigma$ ranges of $-0.73$ to $0.71$ at the wavelength of $1635$~nm and  $-0.74$ to $0.70$ at $1510$~nm (Fig.~\ref{fig:nonlinear_exp}\bsans{c}). Like the theoretical results, these points do not lie in a single plane as we would expect from the experimental results, and reach higher values for circular polarization than those in the linear regime. In the non-resonant regime at $1590$~nm, while $\sigma$ values are non-zero ($-0.39$ to $0.58$), the low output intensity means other fabrication errors were likely to play a large role. The deviation between the experimental and theoretical results is attributed to the same factors as in the linear case. 
Additionally, the THG exhibits greater sensitivity to all possible imperfections and pump parameters compared to the linear regime. Considering the impact of the modes at the THG wavelength, it is challenging to fully replicate the simulated result in a real finite sample.

\section{Conclusions} 

We have studied the effect of resonances on the polarization conversion in chiral dielectric metasurfaces, for both linear and nonlinear regimes. We have demonstrated that metasurfaces composed of a monoclinic lattice of achiral meta-atoms possess a chiral response that can be employed for active polarization engineering. We have verified that such an intrinsic chirality of the metasurface can transform input linearly polarized light into elliptically polarized light, and we have demonstrated that this effect can be used to control the polarization of the generated third harmonic field. We believe that our results provide the first step in exploring polarization transformations in the nonlinear regime for resonant chiral metasurfaces, and they lay the foundation for future work to optimize such phenomena for applications in chiral sensing, chirality encoding, and chiral imaging.

\begin{acknowledgement}
Y.K. thanks M. Gorkunov, K. Konishi, and O. Martin for useful discussions of monoclinic metasurfaces and nonlinear chiral effects. The authors thank S. Xiao for valuable suggestions on the fabrication.
\end{acknowledgement}

\begin{funding}
Australian Research Council (Grant No. DP210101292), International Technology Center Indo-Pacific (ITC IPAC) via Army Research Office (Contract FA520923C0023), National Natural Science Foundation of China (Grants No. 12261131500, No. 12025402, and No. 12334016),
Shenzhen Fundamental Research Projects (Grant No. JCYJ20241202123719025).
\end{funding}

\begin{authorcontributions}
IT, DH, PT contributed equally to this work (marked by $^{\ddagger}$). 

YK and QS conceived the idea and supervised the research.
DH and IT developed the theoretical model and performed numerical simulations. FL fabricated the samples. DH and PT designed and carried out the experiments. IT and PT drafted the initial version of the manuscript.  All authors have accepted responsibility for the entire content of this manuscript and consented to its submission to the journal, reviewed all the results and approved the final version of the manuscript.

\end{authorcontributions}

\begin{conflictofinterest}
Authors state no conflict of interest.
\end{conflictofinterest}

\begin{dataavailabilitystatement}
The datasets generated during and/or analyzed during the current study are available from the corresponding author on reasonable request.

\end{dataavailabilitystatement}

\printbibliography
\onecolumn

\appendix

\section{Numerical simulations}

All numerical simulations were performed in the Wave Optics module of COMSOL Multiphysics.
The near-field distributions, resonant wavelengths, and $Q$-factors are simulated using the eigenfrequency solver.
Linear and nonlinear transmission simulations are simulated in the frequency domain.
The metasurface was placed on a semi-infinite substrate surrounded by a perfectly matched layer mimicking an infinite region in the vertical direction.
The simulation area is the unit cell with Floquet periodic boundary conditions which simulates an infinite size of the metasurface in a transverse plane.
The dispersion of the refractive index of Si is extracted from the ellipsometry data (see SM of Ref.~\cite{Toftul2024PRL}), while that of SiO$_2$ is taken from Refs.~\cite{Malitson1965JOSA,Polyanskiy2024SciData}.

\begin{figure}
	\centering
	\includegraphics[width=0.8\linewidth]{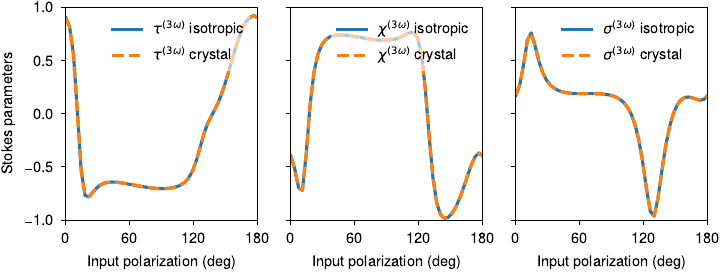}
	\caption{Stokes parameters of the THG signal for the $\lambda_{\text{pump}} = 1555~\text{nm}$ for crystalline Si with $\hat{\chi}^{(3)}$ defined by \eqref{eq:chi3_crystal} and for isotropic Si, for which nonlinear polarization is defined by Eq.~\eqref{eq:chi3_isotropic}.}
	\label{fig:chi3_isotropic_vs_crystal}
\end{figure}

The background field is set manually via custom code using Fresnel equations.
The third harmonic generation (THG) is calculated in the undepleted pump approximation using the domain polarization feature~\cite{COMSOL_SHG_example}. The nonlinear polarization current is calculated as $P_{i}^{(3\omega)} = \varepsilon_0 \hat{\chi}^{(3)}_{ijkm} E_{j}^{(\omega)} E_{k}^{(\omega)} E_{m}^{(\omega)}$, 
where $\hat{\chi}^{(3)}$ tensor has 21 nonzero elements based on the Si symmetry class $m3m$ (227-th space group)~\cite[Table~1.5.4]{boyd2008nonlinear}:
\begin{align}
	\begin{split}
		xxxx &= yyyy = zzzz, \\
		yyzz &= zzyy = zzxx = xxzz = xxyy = yyxx,   \\
		yzyz &= zyzy = zxzx = xzxz = xyxy = yxyx,   \\
		yzzy &= zyyz = zxxz = xzzx = xyyx = yxxy. 
	\end{split}
	\label{eq:chi3_crystal}
\end{align}
Here $ijkm$ is the shorthand for $\hat{\chi}^{(3)}_{ijkm}$ for $i,j,k,m = x,y,z$.
Among 21 nonzero elements only 4 are independent. We assume that the crystallographic axes are aligned with the
metasurface grating direction and incident field direction, i.e. with the base Cartesian unit vectors $(\mathbf{\hat{x}}, \mathbf{\hat{y}}, \mathbf{\hat{z}})$.
While it is possible to find values of each component experimentally in some approximations, their values are of the same order of magnitude~\cite{Zhang20105thInternationalSymposiumonAdvancedO,Moss1989OL}. 
For simplicity we set these to be equal. Furthermore, even approximation of the isotropic Si works well in this particular scenario: 
\begin{equation}
	\vb{P}^{(3\omega)} \approx \varepsilon_0 \chi^{(3)} \left(\vb{E}^{(\omega)} \right)^2 \vb{E}^{(\omega)},
	\label{eq:chi3_isotropic}
\end{equation}
where $\chi^{(3)}$ is a scalar. On Fig.~\ref{fig:chi3_isotropic_vs_crystal} we compare the polarization output of THG signal for the $\lambda_{\text{pump}} = 1555~\text{nm}$. We find that there are no practical difference in the results.

Once the total fields are calculated for the specific background field, $\mathbf{E}_{{\text{bg}}}$, the complex co-polarized transmission amplitude coefficients are calculated as $t^{(n\omega)}_{^{\text{R}} _{\text{L}} } = \braket{\mathbf{\hat{e}}_{^{\text{R}} _{\text{L}}}}{\mathbf{E}_{{\text{bg}}}^{(n\omega)}} =  \frac{1}{A} \iint\limits_{A} \mathbf{\hat{e}}^{*}_{\pm} \cdot \mathbf{E}_{{\text{bg}}}^{(n\omega)} (x,y,z_0) \dd x \dd y$, where $A$ is the area of the $z=z_0$ plane located at the edge of the simulation area from the opposite side of excitation, and $\mathbf{\hat{e}}_{\pm} = (\mathbf{\hat{x}} \pm \iu \mathbf{\hat{y}} )/\sqrt{2}$ are the unit vectors in the circuar polarization basis.
Integration over surface $A$ averages the output signal over the angles, so it gives only the $0$-th diffraction order.
Finally, the transmission coefficients are calculated as $T^{(\omega)}_{^\text{R} _{\text{L}}} = \frac{n_{\text{subs}}}{n_{\text{host}}} \abs{t_{^\text{R} _{\text{L}}}^{(\omega)}}^2$, and the output harmonic intensity is $I_{^\text{R} _{\text{L}}}^{(3\omega)} \propto \abs{t_{^\text{R} _{\text{L}}}^{(3\omega)}}^2$, where the proportionality coefficient is unimportant within the scope of this work.

\section{Sample fabrication}

The silicon metasurface was fabricated with a combined process of electron-beam (E-beam) evaporation, E-beam lithography and inductive coupled plasma (ICP) etching. Basically, $400$~nm silicon film was deposited by E-beam evaporation, and then covered with $26$~nm Cr layer and $80$~nm E-beam resist (PMMA A2). The nanostructures designed in PMMA are patterned via an E-beam aligner (Raith E-line, $30$~kV) and developed in MIBK: IPA for 60 s. Taking the PMMA as a mask, the Cr layer was etched with O$_2$ and Cl$_2$ (Gas flow of O$_2$ is $12.5$~sccm, gas flow of Cl$_2$ is $37.5$~sccm, pressure is $12$~mTorr) in ICP. The Si metasurface was achieved by further etching the Si layer with a mixture of SF$_6$ and C$_4$F$_8$ (Gas flow of SF$_6$ is $35$~sccm, gas flow of C$_4$F$_8$ is $40$~sccm, pressure is $15$~mTorr ) in ICP and removing the Cr layer with chromium etchant.

\section{Optical experiments}

\begin{figure}
	\centering
	\includegraphics[width=0.7\linewidth]{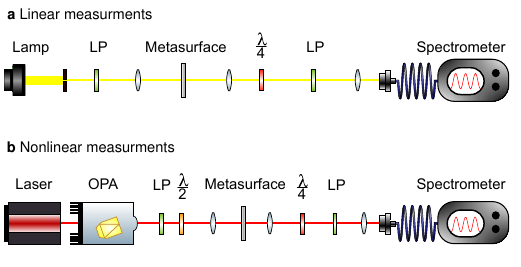}
	\caption{\textbf{Linear and nonlinear experimental setups.} \bsans{a} Linear measurements  were taken with a quartz tungsten-halogen lamp light source. A linear polarizer (LP) was used to convert the unpolarized light to linearly polarized. \bsans{b} Nonlinear measurements were taken from a femtosecond laser. An LP was used to ensure the laser light was fully linearly polarized before the half-waveplate rotated its polarization. Various lenses were used to align the system, focus the light onto the sample and couple it to the fiber spectrometer. }
	\label{fig:experimental_setup}
\end{figure}

For the transmission measurements, a quartz tungsten-halogen lamp was used as the light source (Fig.~\ref{fig:experimental_setup}\bsans{a}). The emitted light passed through a linear polarizer mounted on the motorized rotatable stage enabling precise adjustment of the linear polarization angle. Next, the polarized light is focused onto the sample by a CF$_2$ lens with a focal length of $50$~mm. The transmitted signal after passing through the sample was collected by a 20X Mitutoyo Plan Apo NIR  objective lens with NA=0.4. Then, the light passed through a quarter-wave plate and a linear polarizer, both mounted on rotatable stages, enabling the extraction of all polarization parameters through various combinations of orientations. Subsequently, the light collected was coupled into an optical fiber by an aspherical lens with a focal length of $8$~mm and delivered to the  NIRQuest Ocean Optics spectrometer.

For nonlinear transmission measurement, a laser system comprising a femtosecond laser (Ekspla Femtolux 3) operating at a wavelength of $1030$~nm and an optical parametric amplifier (MIROPA Hotlight Systems) was used as THG pump source (Fig.~\ref{fig:experimental_setup}\bsans{b}). The optical parameter amplifier generated tunable NIR radiation in the range of $1500$--$1700$~nm. The laser pulses had a duration of $250$~fs, a repetition rate of $5.14$~MHz, an average power of $20$~mW and a spot size of diameter $25$~$\upmu$m. To selectively excite specific wavelengths, the laser was tuned in $10$~nm increments across the wavelength range, and the resulting output was measured at each selected wavelength. The NIR laser radiation was linearly polarized with a vertical linear polarizer. Next, the polarization was rotated by a half-wave plate mounted on a rotatable stage. The polarized laser beam was focused onto the sample by a CaF$_2$ lens with a focal length of $50$~mm. The transmitted THG signal was collected after passing through the sample by a 20X Mitutoyo Plan Apo NIR objective lens with NA=0.4. Subsequently, the THG signal was directed through a Thorlabs superachromatic quarter-wave plate (range $325-1100$~nm) and a linear polarizer, both mounted on rotatable stages, allowing for the extraction of all polarization parameters by varying their orientations. Finally, the THG signal was coupled into an optical fiber using an aspherical lens with a focal length of $8$~mm and delivered to the QE Pro Ocean Optics spectrometer.

\section{Third-Harmonic Generation mode}

\begin{figure}[hh]
	\centering
	\includegraphics[width=1\linewidth]{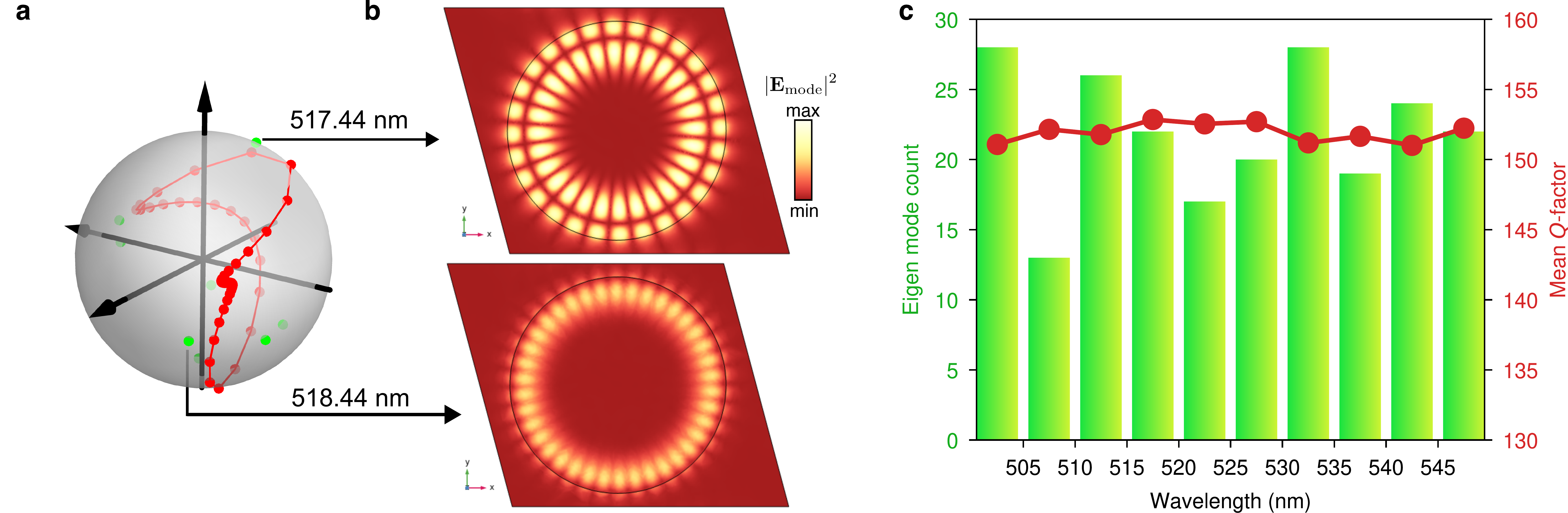}
	\caption{\textbf{Third harmonic mode analyses.} \textbf{a} Poincar\`e sphere for $1555$~nm THG with eigenmodes. The green points correspond to the polarization states of the eigenmodes within $2$~nm of the third harmonic frequency. The red line is the theoretical polarization states of the output. \textbf{b} Examples of mode profiles corresponding to $517.44$~nm and $518.44$~nm. \textbf{c} Histogram of the eigen mode modes with $5$~nm width bin and corresponding mean $Q$-factor. Simulations were done for the $\varepsilon_{\text{SiO}_2} = 2.14$ and $\varepsilon_{\text{Si}} = 21.5 - \iu 0.14$.}
	\label{fig:eigenmode_sphere}
\end{figure}

Fig.~\ref{fig:eigenmode_sphere}\textbf{a} shows the TH output state for the $1555$~nm, and shows that the possible excited resonant modes all lie close to this output path. This indicates that the complex shape of the path is likely a result of different coupling amplitudes to different modes at various points on the sphere. For example, the ``loop'' is positioned in between three different excited modes on the sphere, which likely contribute more than the other modes positioned at other locations. However, the complete analyses of the output THG polarization profile contains a lot of challanges which are outside of the scope of current work. Fig.~\ref{fig:eigenmode_sphere}\textbf{c} shows the mode density in the relevant to the THG spectral region, which indicates a huge number of mode one has to potentially consider.

\begin{figure}[h]
	\centering
	\includegraphics[width=0.75\linewidth]{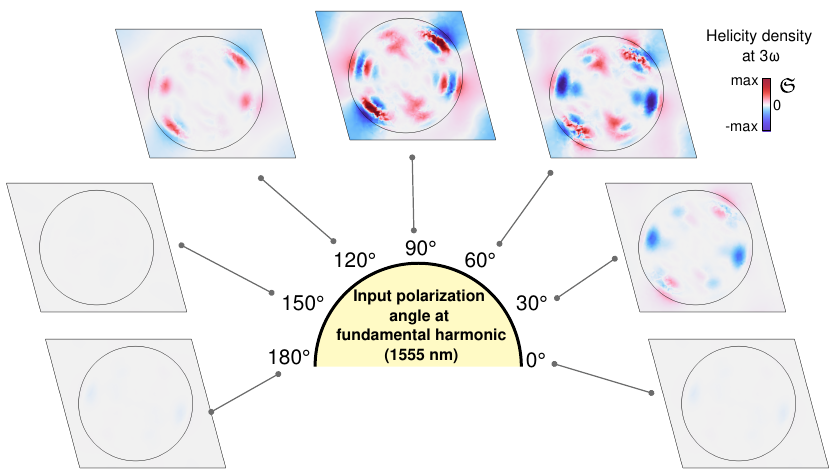}
	\caption{Helicity density at the third harmonic for different input polarization angles. Min and max values of the color bar are of the same scale for each input polarization angle.}
	\label{fig:helicity_3omega}
\end{figure}

Fig.~\ref{fig:helicity_3omega} shows a strong dependence of helicity density at the third harmonic for different input polarization angles. 

\section{Stokes Parameter Graphs for all reported modes}

\begin{figure}[hh]
	\centering
	\includegraphics[width=1\linewidth]{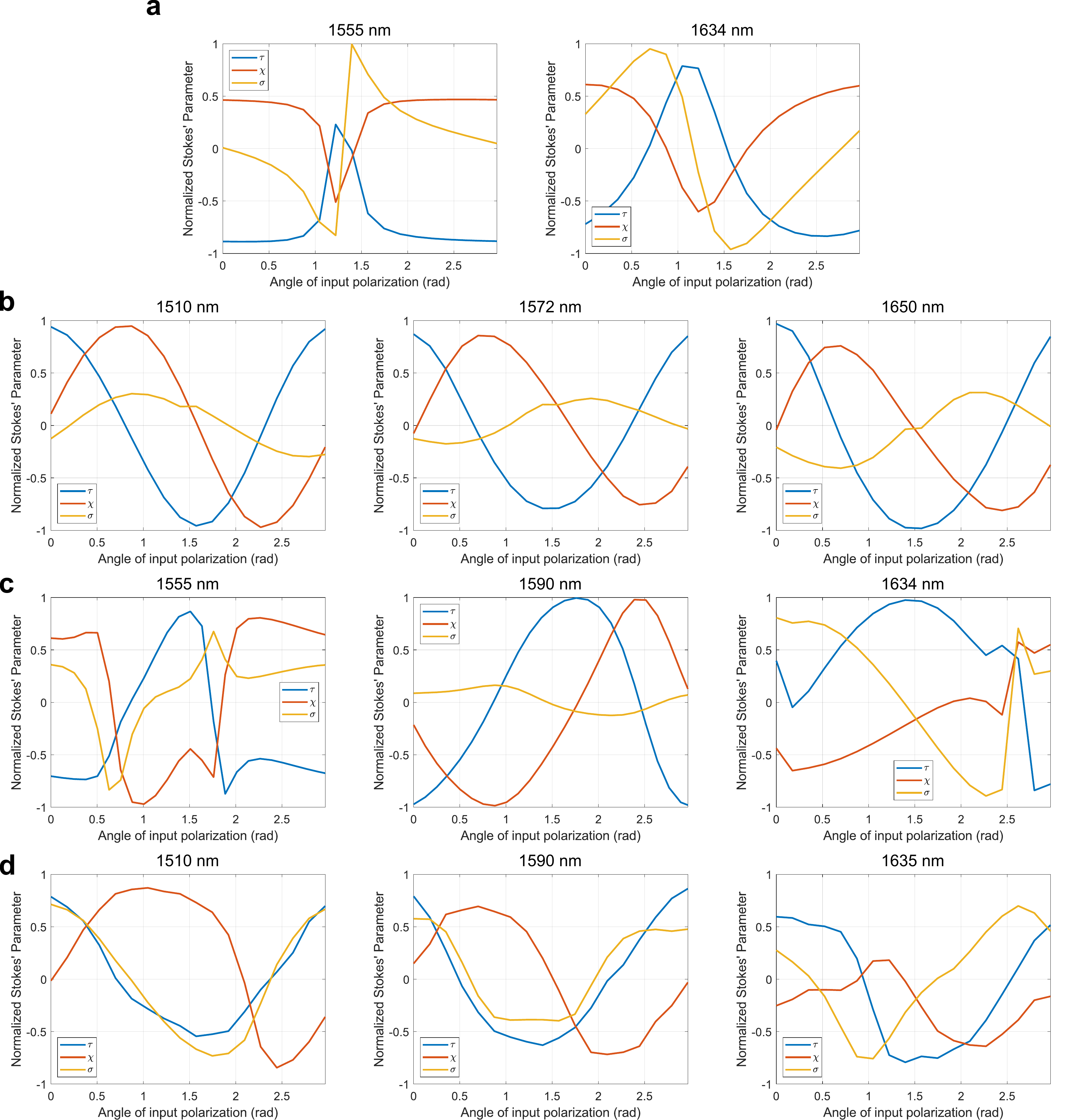}
	\caption{\textbf{Stokes parameters for all examined modes.} Linear \textbf{(a)} theoretical and \textbf{(b)} experimental results. Nonlinear \textbf{(c)} theoretical and \textbf{(d)} experimental results. }
	\label{fig:Stokes}
\end{figure}

Figure~\ref{fig:Stokes} presents the Stokes parameter graphs for all examined modes, comparing theoretical and experimental results. Panels \bsans{a} and \bsans{b} show the Stokes parameters for the linear case, highlighting similarities and differences between theory and experiment. Panels \bsans{c} and \bsans{d} display the corresponding results for the nonlinear case, demonstrating the distinct polarization behavior of THG.

\section{Theoretical and Experimental Heatmaps}

\begin{figure}
	\centering
	\includegraphics[width=0.9\linewidth]{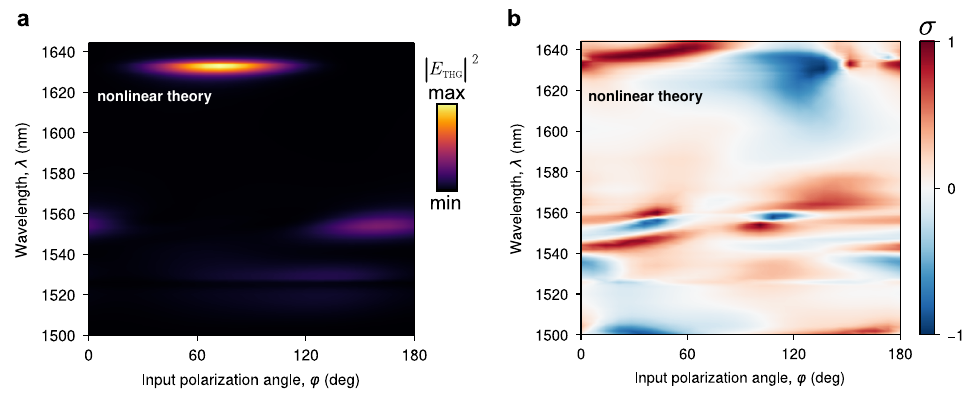}
	\caption{Theoretical \bsans{a} THG efficiency and \bsans{b} circular polarization for wavelength range studied.}
	\label{fig:heatmaps_nonlinear}
\end{figure}

\begin{figure}
	\centering
	\includegraphics[width=0.9\linewidth]{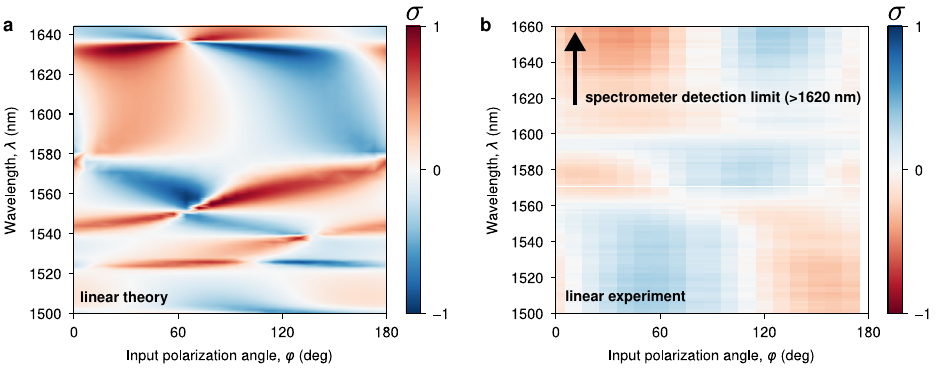}
	\caption{Theoretical \bsans{a} and experimental \bsans{b} circular polarization heatmaps for the for wavelength range studied.}
	\label{fig:heatmaps_linear_with_exp}
\end{figure}

Figures~\ref{fig:heatmaps_linear_with_exp} provides a comparison of theoretical and experimental heatmaps of the output degree of circular polarization. Figure~\ref{fig:heatmaps_nonlinear} shows nonlinear theory of the wide wavelength range.

\end{document}